\documentclass[cameraready]{Interspeech}

\title{Adaptive Oscillatory Inductive Bias for Modeling Sharp Prosodic Dynamics in Diffusion-Based TTS}

\author[orcid=0000-0002-3606-6664]{Sandipan}{Dhar}
\author[orcid=0000-0002-7294-6757]{Nirmesh J.}{Shah}
\author[orcid=0000-0003-3590-503X]{Ashishkumar P.}{Gudmalwar}
\author[orcid=0000-0001-5602-2901]{Pankaj}{Wasnik}

\address{
    Media Analysis, Sony Research India 
}

\email{\{sandipan.dhar,nirmesh.shah,ashishkumar.gudmalwar1,pankaj.wasnik\}@sony.com}


\usepackage{comment}
\usepackage{caption}
\usepackage{cite}
\usepackage{amsmath,amssymb,amsfonts}
\usepackage{algorithmic}
\usepackage{graphicx}
\usepackage{subcaption}
\usepackage{textcomp}
\usepackage{xcolor}
\usepackage{multicol}
\usepackage{multirow}
\usepackage{float}
\usepackage{cite}
\usepackage{url}
\usepackage{amsmath,amssymb}
\usepackage{bm} 
\usepackage{xurl}

\begin{document}

\maketitle

\begin{abstract}
Diffusion-based text-to-speech (TTS) models have achieved significant improvements in speech quality. However, modeling sharp prosodic transitions and rapid pitch variations in expressive speech remains challenging. Existing diffusion-based TTS decoders commonly utilize periodic nonlinearities such as Snake activation function to capture harmonic structures, but this activation funcation provides limited adaptability when modeling abrupt amplitude and frequency variations. In this paper, we investigate the role of oscillatory inductive bias in diffusion-based TTS decoders and introduce an adaptive oscillatory nonlinearity that enables controllable periodic modulation while maintaining signal stability through a linear bypass component. We refer the resulting TTS system as OscillaTTS. Experiments on the LJSpeech and Emotional Speech Dataset show consistent improvements across objective and subjective evaluations, indicating improved modeling of expressive prosodic dynamics. 
\end{abstract}
\section{Introduction} 
Recent advances in deep learning have significantly improved the quality of text-to-speech (TTS) synthesis systems, enabling highly natural and intelligible synthetic speech \cite{xie2025towards,Ref-1-TTSsurvey}. In particular, diffusion-based TTS models have demonstrated strong performance in generating high-fidelity speech by progressively refining acoustic representations \cite{Ref-1-TTSsurvey,Diffusion-survey,Grad-TTS,jeong2021diff,kim2022guided}. Despite these advances, accurately modeling sharp prosodic transitions and rapid pitch variations remains challenging, especially in expressive speech scenarios such as emotional narration or conversational dialogue. These transitions often involve abrupt changes in pitch, energy, and harmonic structure, which remain challenging for current TTS systems to model reliably.
\par
A typical TTS system consists of three stages: extraction of linguistic characteristics, acoustic modeling, and waveform reconstruction using neural vocoders \cite{dutoit1997introduction,gudmalwar2024vecl}. One of the prominent diffusion-based TTS architectures, StyleTTS2 \cite{StyleTTS2}, integrates phoneme-level encoders such as PLBert \cite{PLBert} to capture linguistic context while neural decoders generate the final speech waveform from acoustic representations. While these architectures have improved speech naturalness, the nonlinear transformations inside decoder and vocoder components play an important role in determining how effectively temporal speech dynamics are modeled. In particular, activation functions influence how neural networks represent periodic and aperiodic structures in speech signals.
\par
Speech signals exhibit strong quasi-periodic structures due to vocal fold vibration during voiced speech, resulting in harmonic patterns that evolve over time \cite{quatieri2002discrete}. To better model such oscillatory structures, periodic activation functions have recently been adopted in neural audio models. In particular, the Snake activation function \cite{Snake} employs a periodic inductive bias that enables neural networks to capture harmonic patterns such as pitch and rhythm more effectively for modeling smooth periodic patterns \cite{StyleTTS2,BigVGAN,EzAudio}. However, these activations may struggle to represent rapid prosodic variations that occur in expressive speech, such as abrupt pitch transitions and/or at voiced–unvoiced boundary points. Moreover, many periodic activations rely on fixed-frequency parameters, which can limit their flexibility when modeling diverse speech dynamics across speakers and emotional conditions.
\par
In this work, we investigate the role of oscillatory inductive bias in diffusion-based TTS decoders and introduce an adaptive oscillatory nonlinearity that enables controllable periodic modulation while preserving signal stability. The proposed activation combines periodic structure with a learnable  parameter, allowing the network to dynamically adapt its oscillatory response to rapid changes in speech dynamics. We incorporate this mechanism into the decoder of the StyleTTS2 model and refer to the resulting system as \textit{OscillaTTS}\footnote{Demo \url{https://research.sri-media-analysis.com/interspeech26-oscilla-tts/}}. The proposed activation function is defined as $x + \tanh(\alpha \sin^2(x))$, where the periodic component models oscillatory speech patterns and the learnable parameter $\alpha$ enables adaptive modulation of the activation response, inspired by the HOSC activation \cite{HOSC}. The additional linear component preserves signal stability and helps to maintain the underlying structure of the speech signal during abrupt transitions. This design enables the model to capture both smooth harmonic variations and sharp prosodic changes more effectively than fixed periodic activations.

The main contributions are summarized as follows:

\begin{itemize}
\item We investigate the role of oscillatory inductive bias in diffusion-based TTS decoders for modeling expressive speech dynamics.
\item We propose an adaptive oscillatory activation function that enables controllable periodic modulation while preserving signal stability.
\item We integrate the proposed activation into the StyleTTS2 architecture and demonstrate improved modeling of sharp prosodic transitions.
\item Extensive experiments on the LJSpeech \cite{LJSpeech} and Emotional Speech Dataset (ESD) \cite{ESD} show consistent improvements in objective and subjective evaluation metrics compared to the considered baselines.
\end{itemize}
\begin{figure*}[h!]
    \centering
    \includegraphics[width=0.98\linewidth]{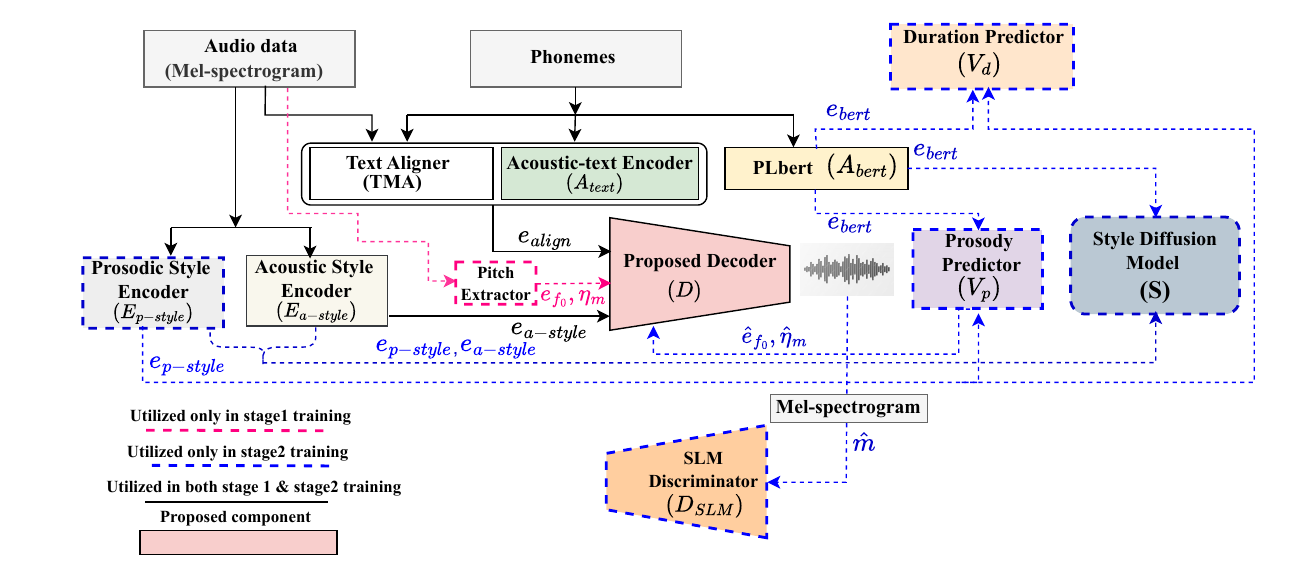}
    \vspace{-0.2cm}
    \caption{Schematic overview of the proposed OscillaTTS architecture based on StyleTTS2.}
    \label{Fig-2:OscillaTTS-model}
    \vspace{-0.5cm}
\end{figure*}
\section{Adaptive Oscillatory Inductive Bias for Diffusion-Based TTS}
\label{Proposed Work}

In this section, We first present the overall architecture and training pipeline, and the integration of the proposed Oscilla activation within the decoder. As illustrated in Fig.~\ref{Fig-2:OscillaTTS-model}, the architectural framework and training mechanism follow the StyleTTS2 architecture \cite{StyleTTS2}. Similar to \cite{StyleTTS2}, the training procedure consists of two stages: a pre-training stage ($stage1$) and a joint training stage ($stage2$). Given phoneme-level inputs $\boldsymbol{p} \in \boldsymbol{P}$ corresponding to textual content and their associated speech features $\boldsymbol{m} \in \boldsymbol{M}$ (mel-spectrograms), we employ a pre-trained acoustic text encoder $A_{\text{text}}$ (a bidirectional LSTM \cite{StyleTTS}) and a pre-trained prosodic text encoder $A_{\text{bert}}$ (PLBert \cite{PLBert}) to extract feature embeddings $\boldsymbol{e_{\text{text}}}$ and $\boldsymbol{e_{\text{bert}}}$, respectively.

The Transferable Monotonic Aligner (TMA) \cite{StyleTTS} is then used to align phoneme-level information with the corresponding acoustic features, producing the speech--phoneme alignment $\boldsymbol{s}_{\text{align}}$. The aligned phoneme representation $\boldsymbol{e}_{\text{align}}$ is obtained through the dot product between $\boldsymbol{e}_{\text{text}}$ and $\boldsymbol{s}_{\text{align}}$. To capture prosodic characteristics, a pre-trained JDC network \cite{StyleTTS} extracts pitch information from $\boldsymbol{m}$, denoted as $\boldsymbol{e}_{f_0}$ where $\boldsymbol{e}_{f_0} = JDC(\boldsymbol{m})$. Additionally, the energy representation $\boldsymbol{\eta}_{m}$ is obtained as the logarithmic norm corresponding to the frame-level energy of $\boldsymbol{m}$. Furthermore, the acoustic style encoder $E_{\text{a-style}}$ and the prosodic style encoder $E_{\text{p-style}}$ extract acoustic and prosodic style embeddings $\boldsymbol{e}_{\text{a-style}}$ and $\boldsymbol{e}_{\text{p-style}}$ from the mel-spectrogram. The prosody predictor $V_{p}$ estimates pitch and energy features (i.e., $\boldsymbol{\hat e}_{f_0}$ and $\boldsymbol{\hat \eta}_m$), while the duration predictor $V_{d}$ predicts phoneme durations conditioned on phoneme embeddings and prosodic style representations (i.e., $\boldsymbol{e}_{\text{bert}}$ and $\boldsymbol{e}_{\text{p-style}}$) \cite{StyleTTS}.

\subsection{Decoder Architecture with Oscilla Activation}
\textbf{Stage 1 training:}
In the proposed model, the decoder $D$ is implemented using the iSTFT-Net vocoder architecture \cite{ISTFTNET}. To introduce an adaptive oscillatory inductive bias for modeling speech dynamics, we integrate the proposed Oscilla activation function within the decoder layers. The Oscilla activation is defined as
\[
x + \tanh(\alpha \sin^2(x)),
\]
where the periodic component $\sin^2(x)$ captures oscillatory structures in speech signals, while the learnable parameter $\alpha$ enables adaptive modulation of the activation response.

The decoder reconstructs the mel-spectrogram $\boldsymbol{\hat m}$ as
\[
\boldsymbol{\hat m} = D(\boldsymbol{e}_{\text{align}}, \boldsymbol{e}_{\text{a-style}}, \boldsymbol{e}_{f_0}, \boldsymbol{\eta}_{m}).
\]

Let the decoder parameters be represented by $\boldsymbol{\theta}$. 
The objective of the stage~1 training is to estimate parameters $\boldsymbol{\theta}^{*}$ that maximize the likelihood of the target mel-spectrogram $\boldsymbol{m}$:
\begin{equation}
\boldsymbol{\theta}^{*}
=
\underset{\boldsymbol{\theta}}{\arg\max}\;
p\left(
\boldsymbol{m}\mid
D_{\boldsymbol{\theta}}
(\boldsymbol{e}_{\text{align}},
\boldsymbol{e}_{\text{a-style}},
\boldsymbol{e}_{f_0},
\boldsymbol{\eta}_{m})
\right).
\label{Eq:1}
\end{equation}
In practice, the optimal parameters are obtained by minimizing the reconstruction loss
\begin{equation}
\mathcal{L}_{rec} =
\mathbb{E}_{m,p}
\left[
\lVert
\mathbf{m} -
D_{\boldsymbol{\theta}}
(\mathbf{e}_{\mathrm{align}},
\mathbf{e}_{\mathrm{a\text{-}style}},
\mathbf{e}_{f_0},
\boldsymbol{\eta}_{m})
\rVert_1
\right].
\label{rec-loss}
\end{equation}
During stage~1 training, all components associated with the decoder are jointly optimized. 
The remaining loss terms follow the same formulation as in StyleTTS~\cite{StyleTTS}.
\begin{figure*}
    \centering
    \begin{subfigure}[b]{0.31\textwidth}
        \centering
        \includegraphics[height=3.65cm, width=5.5cm]{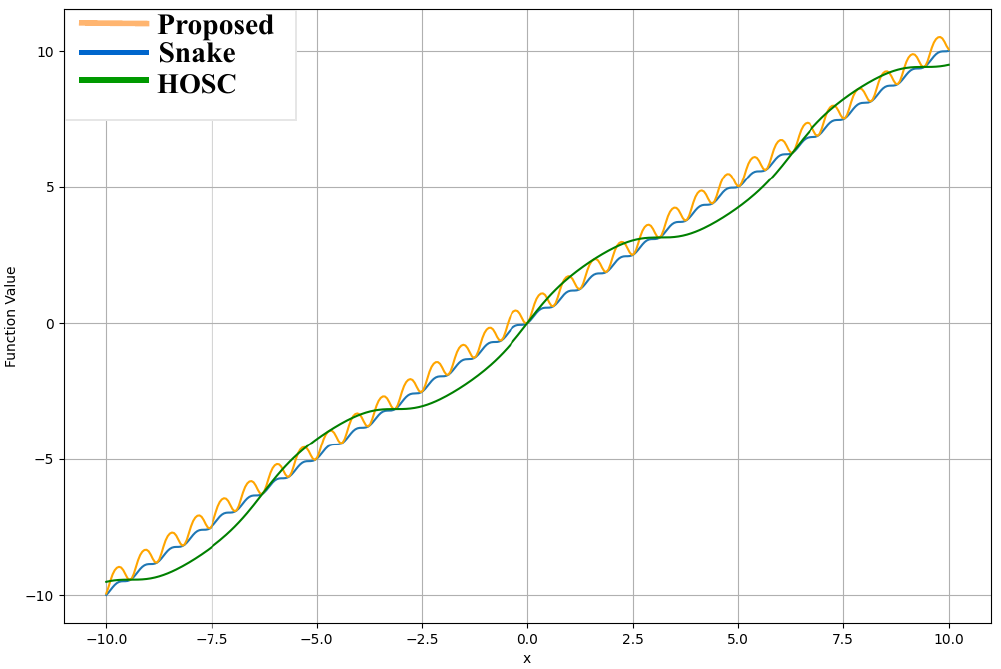}
        \caption{Comparison of Oscilla, Snake, and HOSC activation functions for $a,\alpha$, $\beta$\cite{HOSC}  $\in \{5\}$}
        \label{Fig-1:Activation}
    \end{subfigure}
    \hspace{0.15cm}
    \hfill
    \begin{subfigure}[b]{0.31\textwidth}
        \centering
        \includegraphics[height=3.65cm, width=5.5cm]{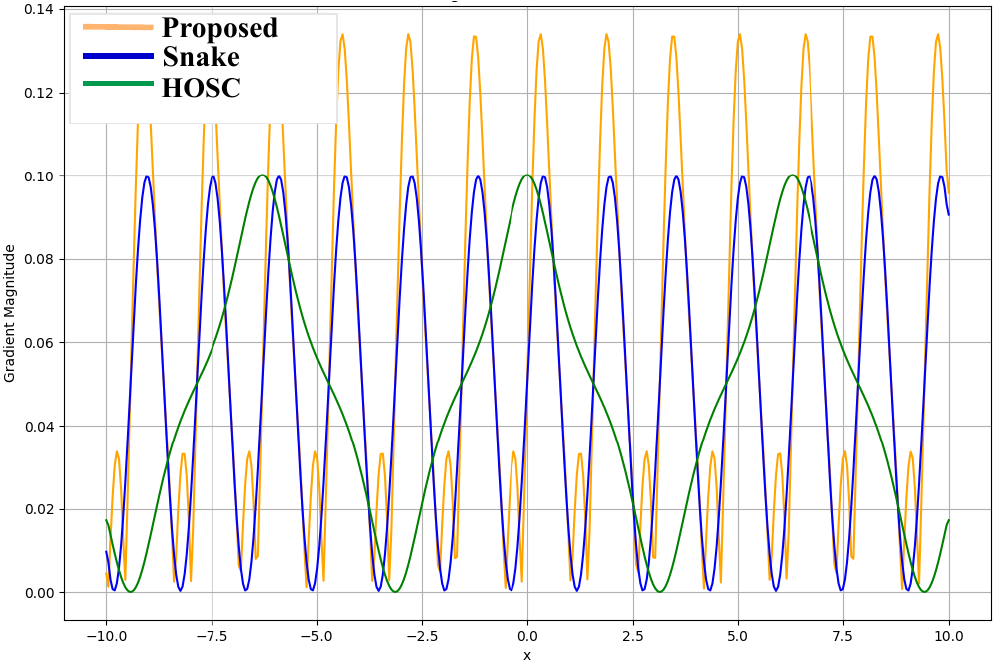}
        \caption{Gradient magnitude comparison of Oscilla, Snake, and HOSC activation functions}
        \label{Fig-2:Gradient-magnitude}
    \end{subfigure}
    \hfill
     \hspace{-0.05cm}
    \begin{subfigure}[b]{0.355\textwidth}
        \centering
        \includegraphics[height=3.65cm, width=5.5cm]
        {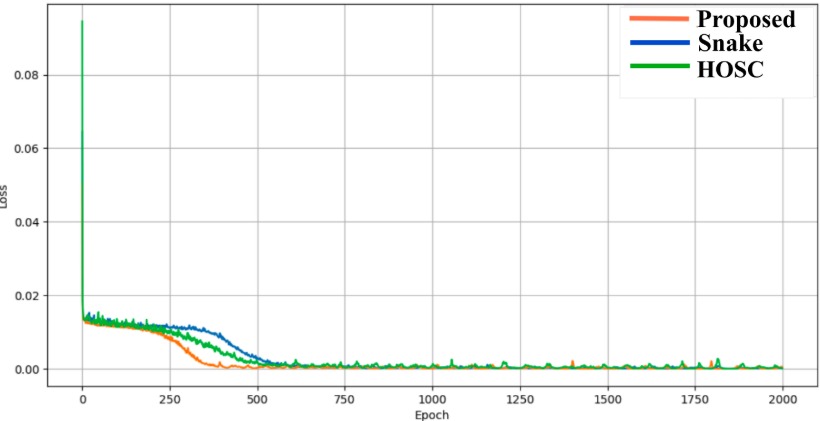}
        \caption{Loss convergence analysis of a neural network using different activation functions}
        \label{Fig-3:Loss}
    \end{subfigure}
    \caption{Comparative analysis of the proposed Oscilla activation with Snake and HOSC. (a) Activation function shapes, (b) gradient magnitude comparison, and (c) loss convergence behavior during neural network training.}
    \label{Fig-4:all}
    \vspace{-0.5cm}
\end{figure*}
\subsection{Analysis of Oscilla Activation for Modeling Sharp Prosodic Transitions}

Speech signals exhibit strong quasi-periodic structures due to vocal fold vibrations during voiced speech \cite{quatieri2002discrete}. Periodic activation functions therefore help neural networks model such oscillatory nonlinearities more effectively. Compared to purely periodic activations such as Snake, incorporating additional nonlinear transformations allows the network to capture richer variations in speech dynamics.




As illustrated in Fig.~\ref{Fig-1:Activation},~\ref{Fig-2:Gradient-magnitude}, Snake \cite{Snake} exhibits a gradient term of the form $\sin(2ax)$, whose amplitude is fixed and whose frequency is controlled by $a$. In contrast, Oscilla produces an oscillatory gradient of the form $\alpha \sin(2x)$ that is modulated by the input-dependent factor $\mathrm{sech}^{2}(\alpha \sin^{2}(x))$. This modulation introduces an implicit gating mechanism: when $\alpha \sin^{2}(x)$ is large, saturation of the $\tanh(\cdot)$ nonlinearity suppresses gradient contributions, whereas smaller values permit stronger oscillatory gradients. Consequently, Oscilla enables adaptive control over oscillatory responses, unlike the fixed-amplitude behavior of Snake. A similar behavior is evident from the Taylor series expansions. In Snake, the parameter $a$ strongly dominates higher-order terms through cubic scaling ($a^{3}$), resulting in fixed-amplitude oscillations that can be difficult to stabilize. In contrast, the parameter $\alpha$ in Oscilla scales higher-order terms linearly and introduces early damping via the $\tanh(\cdot)$ expansion, enabling an adaptive oscillatory response that remains structurally controlled.

To further illustrate the impact of different activations, Fig.~\ref{Fig-3:Loss} shows the convergence behavior of a neural network (four-layer regression model) trained with Snake, HOSC, and Oscilla activations. The training data contains both periodic and aperiodic patterns. The results indicate that Oscilla facilitates stable convergence while maintaining comparable computational complexity to Snake1D. Both activations have $O(n)$ computational complexity, where $n$ denotes the dimensionality of the input feature vector.

\noindent \textbf{Stage 2 training:}

During $stage2$ training, all components shown in Fig.~\ref{Fig-2:OscillaTTS-model} are trained jointly except for the pitch extractor. The style diffusion model $S$ is conditioned on both phoneme embeddings and speaker style embeddings. During inference, the model predicts style embeddings $\boldsymbol{\hat e}_{\text{p-style}}$ and $\boldsymbol{\hat e}_{\text{a-style}}$ from the $\boldsymbol{e}_{\text{bert}}$ representation instead of using the style encoders $E_{\text{p-style}}$ and $E_{\text{a-style}}$, thereby reducing inference time \cite{StyleTTS2}. In addition, a Speech Language Model (SLM) discriminator $D_{SLM}$ \cite{SLM} acts as a critic to evaluate how well the generated mel-spectrogram $\boldsymbol{\hat m}$ preserves acoustic semantic information compared to the original $\boldsymbol{m}$. The remaining loss functions used during $stage2$ training follow the formulation described in \cite{StyleTTS}.

\section{Experimental Setup}
\label{Experimental Details}

We evaluate the proposed approach on two publicly available datasets to assess both standard TTS quality and expressive speech synthesis performance.
LJSpeech \cite{LJSpeech} is used for single-speaker English TTS evaluation. The dataset contains approximately 24 hours of high-quality speech recordings from a single female speaker and is widely used for benchmarking neural TTS systems. To further evaluate expressive speech synthesis and rapid prosodic variations, we conduct experiments on the Emotional Speech Dataset (ESD) \cite{ESD}. The ESD dataset contains recordings from multiple speakers under different emotional conditions. In this work, we focus on the English subset and consider three representative emotions: \textit{Happy}, \textit{Angry}, and \textit{Sad}.

Both datasets were divided into training (80\%), validation (10\%), and testing (10\%) sets. The model is trained using the two-stage training strategy of StyleTTS2 \cite{StyleTTS2}. The pre-training stage ($stage1$) is trained for 200 epochs, while the joint training stage ($stage2$) is trained for 120 epochs. All audio samples are resampled to 24 kHz. We use the AdamW optimizer \cite{Adamw} with $\beta_1=0$, $\beta_2=0.99$, weight decay $\lambda = 10^{-4}$, learning rate $\gamma = 10^{-4}$, and a batch size of 8. All experiments are conducted on a single NVIDIA A100 GPU. The remaining pre-trained components are identical to those used in the baseline StyleTTS2 architecture \cite{StyleTTS2}.
\begin{table}[h!]
\centering
\vspace{-0.2cm}
\caption{Subjective and objective evaluations along with margin of error corresponding to the 95\% Confidence Interval (CI).}
\label{SOTA_Eval}
\setlength{\tabcolsep}{1pt}
\resizebox{0.99\linewidth}{!}{
\begin{tabular}{cccc}
\hline
\multirow{2}{*}{\textbf{Models}} & \textbf{Subjective   } & \multicolumn{2}{c}{\textbf{Objective}}  \\ \cline{2-4} 
& \textbf{Speech Quality $\uparrow$}          & \multicolumn{1}{c}{\textbf{MCD} $\downarrow$}  & \textbf{$F_0$-RMSE $\downarrow$}  \\ \hline
StyleTTS2 \cite{StyleTTS2}           & 81.48   $\pm$ 2.53                         & \multicolumn{1}{c}{6.64 $\pm$ 0.01}          & 0.41 $\pm$ 0.003            \\ 
Proposed   OscillaTTS            & \textbf{86.67 $\pm$ 1.49}                   & \multicolumn{1}{c}{\textbf{6.59 $\pm$ 0.01}} & \textbf{0.35 $\pm$ 0.003}    \\ 
GlowTTS \cite{GlowTTS}                         & 75.79 $\pm$ 2.27                           & \multicolumn{1}{c}{6.85 $\pm$ 0.02}          & 0.4 $\pm$ 0.003             \\ 
GRADTTS \cite{Grad-TTS}            & 83.78 $\pm$ 1.99                           & \multicolumn{1}{c}{6.9 $\pm$ 0.02}           & \textbf{0.35 $\pm$ 0.003}    \\ 
FASTSPEECH2 \cite{FastSpeech-2}                     & 76 $\pm$ 2.77                               & \multicolumn{1}{c}{6.62$\pm$ 0.01}          & \textbf{0.35 $\pm$ 0.003}    \\ \hline
\end{tabular}}
\vspace{-0.5cm}
\end{table}
\begin{table*}[htbp]
\centering
\setlength{\tabcolsep}{2pt}
\caption{Subjective and objective evaluation analysis along with margin of error corresponding to the 95\% CI for expressive speech synthesis task}
\label{Emo}
\vspace{-0.2cm}
\resizebox{0.99\linewidth}{!}{
\begin{tabular}{cccccccccc}
\hline
\multirow{2}{*}{{\textbf{Models}}} & \multicolumn{3}{c}{\textbf{Angry}}                                                           & \multicolumn{3}{c}{\textbf{Happy}}                                                        & \multicolumn{3}{c}{\textbf{Sad}}                                           \\ \cline{2-10} 
 & \multicolumn{1}{c}{\textbf{ES MOS}}$\uparrow$ & \multicolumn{1}{c}{\textbf{MCD}}$\downarrow$ & \textbf{$F_0$-RMSE}$\downarrow$ & \multicolumn{1}{c}{\textbf{ES MOS}}$\uparrow$ & \multicolumn{1}{c}{\textbf{MCD}}$\downarrow$           & \textbf{$F_0$-RMSE}$\downarrow$       & \multicolumn{1}{c}{\textbf{ES MOS}}$\uparrow$ & \multicolumn{1}{c}{\textbf{MCD}}$\downarrow$  & \textbf{$F_0$-RMSE}$\downarrow$ \\ \hline
StyleTTS2                   & \multicolumn{1}{c}{68.8 $\pm$ 2.43}             & \multicolumn{1}{c}{4.68 $\pm$ 0.03}          & 0.67$\pm$ 0.003             & \multicolumn{1}{c}{65.8 $\pm$ 2.52}    & \multicolumn{1}{c}{6.45 $\pm$ 0.03}          & \textbf{0.76$\pm$ 0.003} & \multicolumn{1}{c}{67.34 $\pm$ 2.22}   & \multicolumn{1}{c}{5.4 $\pm$ 0.03}  & 0.5  $\pm$ 0.003   \\ 
Proposed-OscillaTTS                   & \multicolumn{1}{c}{\textbf{70.71 $\pm$ 1.73}}   & \multicolumn{1}{c}{\textbf{4.42$\pm$ 0.03}} & \textbf{0.67 $\pm$ 0.003}    & \multicolumn{1}{c}{\textbf{68.3 $\pm$ 1.93}}    & \multicolumn{1}{c}{\textbf{6.29$\pm$ 0.03}} & 0.77 $\pm$ 0.003          & \multicolumn{1}{c}{\textbf{68.32 $\pm$ 1.56}}   & \multicolumn{1}{c}{\textbf{5.27$\pm$ 0.03}} & \textbf{0.49 $\pm$ 0.004}    \\ \hline
\end{tabular}
}
\vspace{-0.5cm}
\end{table*}
\section{Results and Discussion}
\label{Results & Discussion}
This section evaluates the effectiveness of the proposed adaptive oscillatory inductive bias in diffusion-based TTS.
\subsection{Subjective Evaluation}
To evaluate perceptual speech quality, MUSHRA-style human listening tests were conducted. A total of 25 human subjects (aged between 22 and 36 years old with no reported hearing impairments) participated in the listening tests. Each subject rated synthesized speech samples on a scale of $0-100$, where 100 represents the highest speech quality and 0 represents the lowest. A total of 150 samples were evaluated for each system. Table \ref{SOTA_Eval} compares the proposed method with representative neural TTS baselines. The results show that incorporating the proposed oscillatory inductive bias into the decoder improves perceptual speech quality.
\subsection{Objective Evaluation}
For objective evaluations, we consider Mel Cepstral Distortion (MCD) to measure spectral similarity and $F_0$-RMSE to evaluate pitch modeling accuracy \cite{Berrak_Sisman}. Dynamic Time Warping (DTW) is used to temporally align synthesized and reference speech signals before computing evaluation metrics. We additionally evaluate prosody similarity and intelligibility using AutoPCP and Word Error Rate (WER), respectively. AutoPCP measures utterance-level prosody similarity between synthesized and reference speech, while WER is computed using a Whisper-based speech recognition model \cite{barrault2023seamless}.
\begin{table}[h!]
\centering
\setlength{\tabcolsep}{1pt}
\caption{AutoPCP and WER scores for LJ Speech Dataset}
\label{AutoPCP_WER}
\resizebox{0.99\linewidth}{!}{\begin{tabular}{c|c|c|c|c|c}
     \hline Metrics    & \begin{tabular}[c]{@{}c@{}}Baseline \\ StyleTTS2\end{tabular} & \begin{tabular}[c]{@{}c@{}}Proposed \\ OscillaTTS\end{tabular} & FASTSPEECH2 & GlowTTS & GradTTS \\ \hline
\textbf{AutoPCP} $\uparrow$ & 3.92                                                         & \textbf{4.05}                                                            & 3.94        & 3.67    & 3.91    \\ \hline
\textbf{WER} $\downarrow$     & 2.86                                                         & \textbf{1.85}                                                            & 4.57        & 6.22    & 3.89  \\ \hline
\end{tabular}}
\end{table}

In addition to representative TTS models, we compare the proposed approach with a neural vocoder system such as BigVGAN \cite{BigVGAN}, which also uses periodic activation functions (Snake). The results are shown in Table \ref{BigVGAN}. Overall, the results indicate that incorporating the proposed oscillatory inductive bias improves speech feature reconstruction and prosodic modeling compared to baseline systems.

\begin{table}[h]
\centering
\setlength{\tabcolsep}{2pt}
\caption{Objective Evaluations with BigVGAN Vocoder}
\label{BigVGAN}
\resizebox{0.9\linewidth}{!}{\begin{tabular}{c|c|c|c|c}
     \hline \textbf{Models}           & \textbf{AUTOPCP} $\uparrow$ & \textbf{MCD} $\downarrow$  & \textbf{$F_0$-RMSE $\downarrow$} & \textbf{WER} $\downarrow$  \\\hline
BigVGAN             & 3.87    & 7.56 & \textbf{0.35}    & 7.1  \\
Proposed OscillaTTS & \textbf{4.05}    & \textbf{6.59} & \textbf{0.35} & \textbf{1.85} \\\hline
\end{tabular}}
\vspace{-0.3cm}
\end{table}
\begin{figure}[!htbp]
\centering
\includegraphics[width=0.9\linewidth]{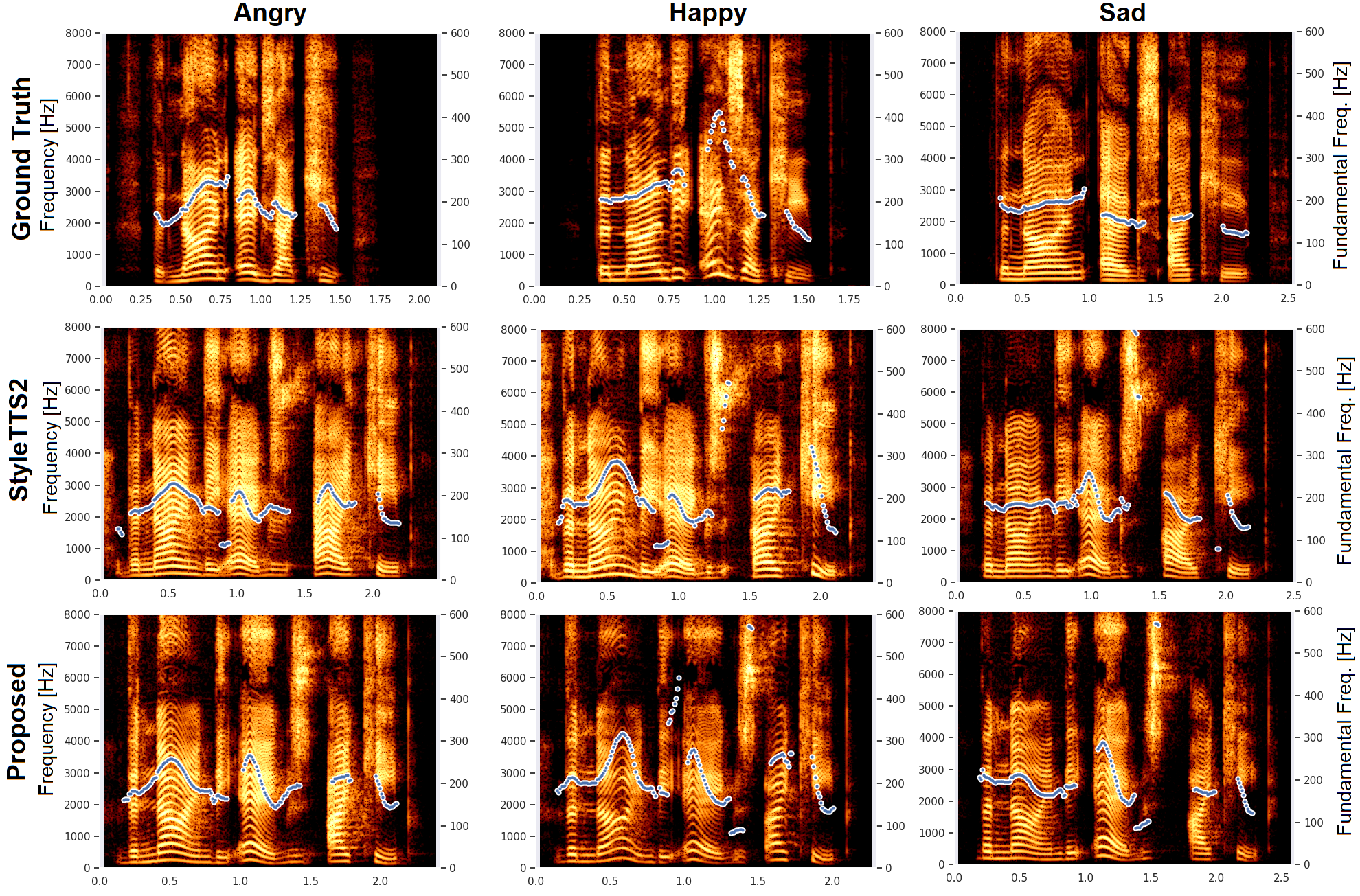}
\vspace{-0.3cm}
\caption{Visualization of Mel-spectrogram along with pitch variation for Angry, Happy and Sad Emotional synthesis.}
\vspace{-0.35cm}
\label{fig:spec}
\end{figure}
\subsection{Expressive Speech Synthesis Evaluation}

To assess the effectiveness of the proposed oscillatory inductive bias for modeling sharp prosodic transitions in expressive speech, we conducted experiments under three emotional speech conditions from the ESD dataset: Angry, Happy, and Sad (Table \ref{Emo}). In addition to speech quality, we conduct subjective emotion similarity (ES) tests where participants rate similarity to the reference emotion on a scale of $0-100$. Objective results on the ESD dataset are further analyzed using AutoPCP and WER (Table \ref{Emo_Autopcp_WER}). The improvements indicate that the proposed inductive bias better captures expressive prosodic dynamics while improving intelligibility.

\begin{table}[h!]
\centering
\setlength{\tabcolsep}{4pt}
\caption{AutoPCP and WER scores for Emotional ESD Dataset}
\label{Emo_Autopcp_WER}
\resizebox{0.88\linewidth}{!}{\begin{tabular}{c|c|c|c|c}
 \hline Metrics                   & Methods             & Angry         & Happy         & Sad           \\ \hline
\multirow{2}{*}{\textbf{AutoPCP}$\uparrow$} & Baseline StyleTTS2  & 3.03          & 3.17          & 2.97          \\
                         & Proposed OscillaTTS & \textbf{3.23} & \textbf{3.21} & \textbf{3}    \\ \hline
\multirow{2}{*}{\textbf{WER}$\downarrow$}     & Baseline StyleTTS2  & 9.21          & 13.3          & 9.72          \\
                         & Proposed OscillaTTS & \textbf{4.05} & \textbf{7.93} & \textbf{7.89} \\ \hline
\end{tabular}}
\vspace{-0.35cm}
\end{table}

Figure \ref{fig:spec} presents a comparative spectrogram analysis for Angry, Happy, and Sad emotional synthesis. Compared to the baseline, the proposed model produces more accurate harmonic structures and more stable pitch trajectories, particularly in regions with rapid prosodic transitions. These observations support that incorporating oscillatory inductive bias improves the modeling of expressive prosodic dynamics.

\subsection{Ablation Study}

To further analyze the contribution of the proposed oscillatory inductive bias, we conduct an ablation study by replacing the Oscilla activation with several alternative activation functions within the same architecture. This experiment helps isolate the impact of the proposed activation on prosodic modeling and speech synthesis quality. In particular, we evaluate variants including Snake1D, ReLU, $\tanh$, $x+\sin(x)$, and $\tanh(\sin(x))$. In addition, we include a variant of Oscilla with a fixed parameter $\alpha$ to examine the importance of adaptive periodic modulation. All experiments are conducted using the LJSpeech dataset while keeping the remaining configuration identical. The results are summarized in Table \ref{Table:Activation}. The proposed Oscilla activation consistently achieves the best performance in terms of both MCD and $F_0$-RMSE. These results indicate that the adaptive oscillatory inductive bias enables the model to better capture harmonic structures and rapid prosodic transitions.
\begin{table}[h]
\vspace{-0.3cm}
\centering
\setlength{\tabcolsep}{2pt}
\caption{Ablation study w.r.t. different activation functions in the proposed OscillaTTS}
\label{Table:Activation}
\vspace{-0.2cm}
\resizebox{0.99\linewidth}{!}{
\begin{tabular}{ccc}
\hline
\textbf{Activation Function}   & \textbf{MCD} $\downarrow$ & \textbf{$F_0$-RMSE} $\downarrow$\\ \hline
Proposed Oscilla (learnable $\alpha$)  & \textbf{6.59 $\pm$ 0.01}  & \textbf{0.35 $\pm$ 0.003}    \\ 
Oscilla (fixed $\alpha=1$) & 6.63 $\pm$ 0.01 & 0.39 $\pm$ 0.003 \\
Snake1D    & 6.64 $\pm$ 0.01 & 0.41 $\pm$ 0.003   \\ 
ReLU       & 8.14  $\pm$ 0.02 & 0.44 $\pm$ 0.003   \\ 
$\tanh$       & 7.87  $\pm$ 0.02 & 0.68 $\pm$ 0.003   \\ 
$x+\sin(x)$   & 12.63 $\pm$ 0.03 & 0.8  $\pm$ 0.003   \\ 
$\tanh(\sin(x))$ & 8.14  $\pm$ 0.02 & 2.56 $\pm$ 0.004   \\ \hline
\end{tabular}
}
\vspace{-0.4cm}
\end{table}

\section{Conclusion}
This work investigated the role of oscillatory inductive bias in diffusion-based TTS systems for modeling expressive speech dynamics. We proposed an adaptive oscillatory activation mechanism that enables controllable periodic modulation while maintaining signal stability. The activation was incorporated into the decoder of the StyleTTS2 architecture, resulting in the OscillaTTS system. Experimental evaluations on the LJSpeech and Emotional Speech Dataset demonstrated consistent improvements across both subjective and objective metrics. These results indicate that incorporating oscillatory inductive bias can improve the modeling of rapid prosodic variations in expressive speech. Future work will explore extending this approach to multi-speaker expressive TTS and singing voice synthesis.
\section{Generative AI Use Disclosure}
Generative AI tools were used solely for language refinement, grammar correction, and improving the overall clarity and readability of the manuscript. These tools assisted in polishing and structuring the text but were not used to generate or design the core research ideas, methodology, experimental setup, analysis, or conclusions presented in this work. All scientific contributions, technical implementations, and interpretations were developed and validated by the authors. In accordance with policy guidelines, no generative AI system is listed as a co-author, and all authors take full responsibility and accountability for the content and integrity of this paper.
\bibliographystyle{IEEEtran}
\bibliography{mybib}
\end{document}